\documentclass[10pt,conference,twoside]{IEEEtran}
\usepackage{graphicx}

\usepackage{epsfig}
\usepackage{graphics}
\usepackage{cite}
\usepackage{amsmath}
\usepackage{amssymb}
\usepackage{amsfonts}
\usepackage{setspace}
\usepackage{color}
\usepackage{setspace}
\usepackage{multirow}
\usepackage{tablefootnote}

\usepackage[skip=2pt,font=small]{caption}

\usepackage{subcaption}
\usepackage{bm, amsmath, amssymb}
\usepackage{amsfonts}
\usepackage {amssymb,amsmath}
\usepackage{cite}
\usepackage{epstopdf}
\usepackage{booktabs}
\usepackage{float}
\usepackage{cite}
\graphicspath{ {SARNOFF_images/} }
\DeclareGraphicsExtensions{.pdf,.png,.jpg}

\DeclareMathOperator{\E}{\mathbb{E}}
\DeclareMathOperator{\D}{\textit{d}}

\IEEEoverridecommandlockouts

\usepackage{tikz}
\usepackage{textcomp}
\usepackage{hyperref}
\usepackage{lipsum}

\newcommand\copyrighttext{%
	\footnotesize "This work has been submitted to the IEEE Sarnoff 2016 conference for possible publication. Copyright may be transferred without prior notice, after which this version may no longer be accessible"}
\newcommand\copyrightnotice{%
	\begin{tikzpicture}[remember picture,overlay]
	\node[anchor=south,yshift=760pt] at (current page.south) {{\parbox{\dimexpr\textwidth-\fboxsep-\fboxrule\relax}{\copyrighttext}}};
	\end{tikzpicture}%
}

\begin{document}
	
	\title{Rate Performance of Adaptive Link Selection in Buffer-Aided Cognitive Relay Networks}
%	\author{Bhupendra Kumar,~\IEEEmembership{Student Member,~IEEE,}
%		and~Shankar~Prakriya,~\IEEEmembership{Senior Member,~IEEE}% <-this % stops a space
%	}
	
	\author{
		\IEEEauthorblockN{Bhupendra Kumar$^{1}$ and Shankar Prakriya$^{2}$\\
			$^{1,2}$ Bharti School of Telecom and Management,
			$^{2}$ Department of Electrical Engineering\\
			Indian Institute of Technology, Delhi, India\\
			E-mail: bkumar0810@gmail.com, shankar@ee.iitd.ac.in}
		%\IEEEauthorblockA{Bharti School of Telecom and Management\\}
	%	\and
		%\IEEEauthorblockN{Prof. Shankar Prakriya,}
	%	\IEEEauthorblockA{
			%Department of Electrical Engineering\\
			%Indian Institute of Technology Delhi \\
		}
	%}

	% avoiding spaces at the end of the author lines is not a problem with
	% conference papers because we don't use \thanks or \IEEEmembership
	\maketitle
	\copyrightnotice

	\begin{abstract}
		
			We investigate the performance of a  two-hop cognitive relay network with a buffered decode and forward (DF) relay. We derive expressions for the rate performance of an adaptive link selection-based buffered relay (ALSBR) scheme with peak power and peak interference constraints on the secondary nodes, and compare its performance with that of conventional unbuffered relay (CUBR) and conventional buffered relay (CBR) schemes. Use of buffered relays with adaptive link selection is shown to be particularly advantageous in underlay cognitive radio networks. The insights developed are of significance to system designers since cognitive radio frameworks are being explored for use in 5G systems.  Computer simulation results are presented to demonstrate accuracy of the derived expressions.
			
	\end{abstract}

	\section{Introduction}	
	\par It has been established that cognitive radio \cite{Haykin2005}, in which  an unlicensed (secondary) user shares the spectrum of the licensed (primary) user, has great potential for alleviating spectrum scarcity.  In particular, underlay cognitive radio networks, in which the secondary node transmits with power that is controlled carefully to ensure that the interference caused to the primary receiver is below an interference temperature threshold, has attracted great research interest \cite{Goldsmith2009}. However, the severe  interference constraints imposed by the primary networks seriously limits the transmit powers, and thereby the rates that can be achieved in the secondary networks.

	\par  The advantages of using of a buffer equipped relay has been demonstrated \cite{Madsen2005}. In non-cognitive two-hop networks, using a buffered-relay, Madsen \cite{Madsen2005} demonstrated rate enhancement in fading channels by averaging the instantaneous rate over multiple time-slots for both the hops. Unlike \cite{Madsen2005}, Bing \cite{Bing2008} utilized two-hops of equal duration, so that the rate was limited by the weaker link. Recently, it has been demonstrated \cite {Zlatanov2011, Zlatanov2013_2, Zlatanov2013_1}  that buffering with adaptive link selection, where either the source-relay or relay-destination link is judiciously selected for transmission, can harness a diversity of two with fixed-rate transmission, and increase the average rate by a factor of two as compared to a conventional buffered relay scheme with adaptive rate. Symbol error rate (SER) performance of such systems is analyzed in  \cite{Islam2014}.  Intuitively, since the sources in cognitive radio networks are power-limited, the use of relays is well motivated. Also, all the  techniques employed to improve performance of relays  can be utilized \cite{Kim2013}.  %{\em We emphasize that all these works are restricted to the non-cognitive context}. 
	\par In \cite{Darabi2014}, an interference cancellation-based scheme is proposed where the primary and the secondary sources  pick one buffer-aided relay each for two-hop  transmission, and address power allocation issues.  In \cite{Darabi2015}, a throughput-optimal adaptive link selection policy is proposed for the secondary two-hop network.   For underlay two-hop buffer-aided relay networks,  a sub-optimal relay selection scheme is proposed in \cite{Chen2014}, and its outage performance is analyzed assuming {\em only} the peak interference constraint (ignoring the peak power constraint). In \cite{Shaqfeh2015}, an {\em overlay} secondary source maximizes its own rate in a link without relays, while assisting the primary to attain its target rate using causal knowledge of the primary message. 
	\par In this paper, assuming  peak interference and peak power constraints on the secondary nodes, we develop closed-form analytical expressions for rate performance of a two-hop {\em underlay} network with a buffered relay.  We compare rate performance of the adaptive link selection scheme with that of conventional buffered and unbuffered relays. To facilitate rate analysis, we  first derive expressions for the joint complementary cumulative distribution function (CCDF) of the link selection parameter and the instantaneous SNR of the selected link. We demonstrate that buffering with adaptive link selection is most beneficial in severely power constrained scenarios typically encountered in underlay cognitive radio. Intuitively, this is because the interference constraints make the transmit power of the source and relay in the two-hop network random variables. This  increases the variance of the SNRs of the two hops, which makes use of a buffer at the relay {\em more important than in cooperative links}.  Since use of the cognitive paradigm is being explored for use in 5G systems, performance of link-level two-hop cognitive radio networks is of great interest to researchers and system designers\cite{Hong2014}\cite{Haider2015}. This this paper, we restrict out attention to rate performance. In the longer version of this paper, we address symbol error rate and delay performance issues.

	\vspace{-.0cm}\section{System Model}\label{sec:SysMod}
	\par We consider a two-hop underlay cognitive network  as depicted in Fig.\ref{fig:sysmod1}. The primary  network consists of  the primary destination (PD), and the secondary or unlicensed network consists of the secondary source (SS), the secondary destination (SD), and a half-duplex (HD) decode and forward (DF) secondary relay (SR). It is assumed that SR is equipped with a buffer. All secondary nodes are assumed to possess a single antenna. The SS-SD direct link is heavily shadowed, necessitating the use of a relay. All channels  between nodes in this network are assumed to be quasi-static, and do not change in the signalling interval, though they change independently from slot to slot.   The channel coefficients of the SS-SR and SR-SD links in a time-slot are denoted by $h_{s}$ and $h_{r}$ respectively, with $h_{i}\sim {\cal CN}(0,\Omega_{h_{i}}), i\in\{s,r\}$. The channel coefficients of the  SS-PD and SR-PD  interference links are denoted by  $g_{s}$ and $g_{r}$ respectively, with $g_{i}\sim {\cal CN}(0,\Omega_{g_{i}}), i\in\{s,r\}$. Let $d_{sr}$, $d_{rd}$, $d_{sp}$ and $d_{rp}$ denote the SS-SR, SR-SD, SS-PD and SR-PD distances respectively. With a path-loss Rayleigh fading channel model, it is clear that  $\Omega_{h_{i}}=d_{ij}^{-\alpha}$ and $\Omega_{g_{i}}=d_{{ip}}^{-\alpha}$ respectively, where $i\in\{s,r\}, j\in\{r,d\}$, and $\alpha$ is  the path-loss exponent.  We also assume zero-mean additive white Gaussian noise of $N_{o}$ variance at all terminals. 
	\par  Underlay CR nodes \cite{Lee2011} use an interference constraint so that SS and SR restrict their instantaneous transmit power in order to limit the peak interference to PD below an interference temperature limit (ITL) ${\cal I}_{p}$. We assume that maximum transmit power  at SS and SR is limited to  $P_{max}$ (peak-power constraint), and define the system SNR as
	 $\gamma_{max}= P_{max}/N_{o}$. With peak interference  and peak power constraints, the instantaneous SNRs $\gamma_{i}$ are given by\cite{Lee2011}:
	\begin{figure}[]
		%	\hspace{6cm}
		\begin{center}
			\includegraphics[width=69mm]{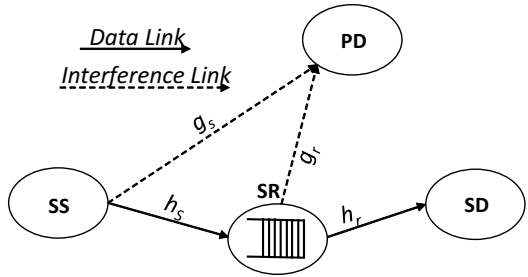}
			\caption{Three Node cognitive buffer-aided relay network.}
			\label{fig:sysmod1}
		\end{center}
		\vspace{-0.75cm}
	\end{figure}
	\begin{eqnarray}\label{eqn:InsSNR}
	\gamma_{i} &=& \min\left\{\gamma_{max},\dfrac{\gamma_p}{|g_{i}|^2}\right\}|h_{i}|^2,\quad  i\in\{s,r\}\label{eqn:link_snrs}
	\end{eqnarray}
	where $\gamma_{p}={\cal{I}}_{p}/N_{o}$. 
	The instantaneous capacity of the two hops is defined as $C_{i}=\log_{2}(1+\gamma_{i})$ $i=s,r$. In the low SNR regime referred to as the peak transmit power regime (PTPR),  $\gamma_{max}$ is small (which ensures that $\gamma_{max}<<\gamma_p/|g_{i}|^{2},$) so that the link SNRs are determined solely by the peak power (and modelled as a exponential random variables). In the high SNR regime referred to as the peak interference power regime (PIPR),  $\gamma_{max}>> \gamma_p/|g_{i}|^{2},$  so that the link SNRs are limited by the interference (and modeled as a ratio of exponential random variables). 
	The probability $p_{i}$, $i\in \{s,r\}$ that the peak interference ($P_{max}|g_{i}|^{2}$) at PD with peak transmit power is higher than ${\cal I}_{p}$ is given by:
	\begin{eqnarray}
	p_{i}&=&\Pr\left\{\gamma_{max}>\dfrac{\gamma_{p}}{|g_{i}|^2}\right\}= e^{-{\mu_{i}}/{\lambda_{i}}}, \quad  i\in\{s,r\}
	\end{eqnarray}
	where $\lambda_{i}=\gamma_{max}\,\Omega_{h_{i}}$ and $\mu_{i}=\dfrac{\gamma_{p}\Omega_{h_{i}}}{\Omega_{g_{i}}}$ are the average SNRs when the SS/SR transmits with 
	$P_{max}$  and $\dfrac{\mathcal{I}_p}{\Omega_{g_{i}}}$ power respectively. Note that when the ratio $\dfrac{\mu_{i}}{\lambda_{i}}$ is $\infty$ and $0$, the corresponding probabilities $p_{i}$ are $0$ and $1$ respectively. Hence $p_i=0$ ($p_i=1$) indicates that the node $i\in\{s,r\}$ is operating in the PTPR (PIPR).
	\par In literature \cite{Lee2011}, expressions have been derived for CDF of the SNRs $\gamma_{i}$ of (\ref{eqn:link_snrs}) of the two links. We will find it convenient to write the CCDF and PDF of link SNRs $\gamma_{i}$  in terms of $p_{i}$ as follows:
	\begin{eqnarray}\label{eqn:F_Gi_s}
	\hspace{-0.2in}F_{\gamma_{i}}(s) \hspace{-0.1in}&=&\hspace{-0.1in} 1-e^{-{s}/\lambda_{i}}\left[1-p_{i}\left(1- \dfrac{\mu_{i}}{s+\mu_{i}}\right) \right],\label{eqn:cdf11}\\
	\hspace{-0.2in}\hspace{-1in}f_{\gamma_{i}}(s) \hspace{-0.1in}&=&\hspace{-0.1in} \dfrac{e^{-{s}/\lambda_{i}}}{\lambda_{i}}\left[1-p_{i}\left(1- \dfrac{\mu_{i}}{s+\mu_{i}}- \dfrac{\lambda_{i}\mu_{i}}{(s+\mu_{i})^2}\right) \right].\label{eqn:f_Gi_s}
	\end{eqnarray}
	In the above, use of $p_{i}=0$ and $p_{i}=1$ results in expressions for PTPR and the PIPR cases respectively.
	\vspace{-0.4cm}\section{Relay Scheme}\label{sec:RelaySchemes}
	\par We assume that both SS and SR have required CSI. Further, they use adaptive modulation  to transmit with maximum rate (the instantaneous capacity of the channel). For rate enhancement, we incorporate the dual-hop adaptive link selection relay scheme of \cite{Zlatanov2011},\cite{Zlatanov2013_1}, where the SS-SR or SR-SD link, whichever has higher capacity, is chosen so as to attain good performance (while ensuring buffer stability).  For analytical tractability (as in  \cite{Zlatanov2011},\cite{Zlatanov2013_1}), the  suboptimal decision function based on the ratio of instantaneous SNRs $\gamma_{s}$ (SS-SR link) and $\gamma_{r}$ (SR-RD link) is used:
	\begin{equation}\label{eqn:linkcrit}
	d = \left\{ 
	\begin{array}{l l}
	1 & \quad \text{if $\dfrac{\gamma_{r}}{\gamma_{s}}\geq$}  \rho \  (\text{SR transmits})\\
	0 & \quad \text{otherwise (SS transmits)},
	\end{array} \right.
	\end{equation}
	where $d$ is the one-bit link-selection parameter, and $\rho$ is a positive statistical parameter that depends on average channel gains. $\rho$  is chosen to maximize  rate while ensuring buffer stability. We assume that SS always has data to transmit, and that SR has an infinite-sized buffer and choose $\rho$  such that the rate is maximised. Hence the average rate of ALSBR is:
	\begin{align}
	\overline{{\mathcal{R}}}^{ALSBR}=\mathbf{\E}_{\gamma_{s},\gamma_{r}}[d C_{r}]=\mathbf{\E}_{\gamma_{s},\gamma_{r}}[(1-d)C_{s}].\label{eqn:rate_expr}
	\end{align}
	 where $\mathbf{\E}_{X}\{.\}$ denotes expectation over the variable $X$. To choose a link, the ALSBR scheme needs the instantaneous SNRs of the two links together with some average channel gains. We assume that  the relay node (using transmitted pilots), performs this selection, and communicates the same to SS prior to signalling in each time-slot.
	 \par In this paper, we compare performance of the ALSBR scheme with  the conventional unbuffered relay (CUBR), which holds the single packet in unit length buffer before relaying it in the next time-slot.  The average rate of CUBR is given by\cite{Madsen2005}:
	 \begin{eqnarray}
	 	\begin{array}{lll}\label{eqn:rate_expr_CUBR}
	 	\overline{{\mathcal{R}}}^{CUBR}=1/2\,\mathbf{\E}_{\gamma_{s},\gamma_{r}}[\min\{C_{s},C_{r}\}].
	 	\end{array}
	\end{eqnarray}
	 In the conventional buffered relay (CBR) scheme that we also use for comparison, the data is stored (hence averaged) for multiple slots before relaying to SD. These slots are equal for SS-SR and SR-SD links, which ensures that the average rate of CBR is \cite{Bing2008}:
	 \begin{eqnarray}
	 \begin{array}{lll}\label{eqn:rate_expr_CBR}
	 \overline{{\mathcal{R}}}^{CBR}=1/2\,\min\{\mathbf{\E}_{\gamma_{s}}[C_{s}],\mathbf{\E}_{\gamma_{r}}[C_{r}]\}.
	 \end{array}
	 \end{eqnarray} 
	\section{Performance Analysis}\label{sec:PerAna}
	\par The CCDF $F^c_{d,\gamma_{s}}(0,x)$ of instantaneous SNR $\gamma_{s}$ of SS-SR link  with link selection parameter $d$, when SS-SR link is selected is:
	\begin{eqnarray*}
		\begin{array}{lll}
			\hspace{-0.0in}F^c_{d,\gamma_{s}}(0,x)\hspace{-0.1 in}&=& F^c_{\gamma_{s}}(x)-F^c_{d,\gamma_{s}}(1,x)\\
			&&\hspace{-0.63cm}=\
			\int\limits_{x}^{\infty}f_{\gamma_{s}}(s) {\D}s-\int\limits_{x}^{\infty}F^c_{\gamma_{r}}(\rho s)f_{\gamma_{s}}(s) {\D}s,
		\end{array}
	\end{eqnarray*}
	where we have used the fact that $F^c_{d,\gamma_{s}}(1,x)=\Pr\{\frac{\gamma_r}{\rho}>\gamma_s>x\}$.  Using (\ref{eqn:F_Gi_s}) and (\ref{eqn:f_Gi_s}) and after some manipulations, it is shown in the Appendix A that $F^c_{d,\gamma_{s}}(0,x)$ is given by (\ref{eqn:CCDF_SR}) in Table~\ref{tab:table_I}, depending on whether $\mu_r=\rho\mu_s$ or $\mu_r\neq\rho\mu_s$.  We define $\lambda_{\rho}$ as the harmonic mean of $\rho\lambda_{s}$ and $\lambda_{r}$ for ease of exposition i.e.  $1/\lambda_{\rho}=1/(\rho\lambda_{s})+1/\lambda_{r}$.
	%Since CCDFs of $\gamma_{s}$ and $\gamma_{r}$ have the %same structure, and
	Similarly:
	\begin{eqnarray*}
		\begin{array}{lll}
			\hspace{-0.0in}F^c_{d,\gamma_{r}}(1,x)\hspace{-0.1 in}&=& F^c_{\gamma_{r}}(x)-F^c_{d,\gamma_{r}}(0,x)\\
			&&\hspace{-0.7cm}=\
			\int\limits_{x}^{\infty}f_{\gamma_{r}}(r) {\D}r-\int\limits_{x}^{\infty}F^c_{\gamma_{s}}(r/\rho )f_{\gamma_{r}}(r) {\D}r.
		\end{array}
	\end{eqnarray*}
	We omit the expression for $F^c_{d,\gamma_{r}}(1,x)$ due to space constraints. It can be shown along similar lines that 
$F^c_{d,\gamma_{r}}(1,x)$ can be obtained from $F^c_{d,\gamma_{s}}(0,x)$ by exchanging the position of SS and SD hence exchanging variables as follows:
\begin{eqnarray}
\begin{array}{llll}
 p_{s}\leftrightarrow p_{r}, & \lambda_{s}\leftrightarrow\lambda_{r}, & \mu_{s}\leftrightarrow\mu_{r},&  \rho\rightarrow1/\rho.
 \end{array}\label{eqn:exchange}
\end{eqnarray}
Consequently,  $\lambda_{\rho}/{\rho}\rightarrow\lambda_{\rho}$.\\
For brevity, we first define an integral $\mathcal{I}_{n}(\mu,\lambda;x)$ 
as follows:
\begin{eqnarray}\label{eqn:I_n_mu_lambda_x}
	\nonumber
	\hspace{-0.4in}\mathcal{I}_{n}(\mu,\lambda;x)\hspace{-0.1in}&=&\int\limits_{x}^{\infty}\dfrac{\mu^{n-1}}{(s+\mu)^n}e^{-{s}/\lambda} {\D}s, \quad n\geq 1\\
	&&\hspace{-0.6cm}=\ \left(\dfrac{\mu}{x+\mu}\right)^{n-1} \exp\left(\dfrac{\mu}{\lambda}\right) E_{n}\left(\dfrac{x+\mu}{\lambda}\right),
\end{eqnarray}
where $E_{n}(x)=\displaystyle\int_{1}^{\infty}\dfrac{e^{-xt}}{t^{n}}dt$ is the generalized exponential integral.
Further, we define integral $\mathcal{I}_{n}(\mu,\lambda)$ for rate as:
\begin{eqnarray}\label{eqn:I_n_mu_lambda}
\hspace{-0.48in}\mathcal{I}_{n}(\mu,\lambda)\hspace{-0.1in}&=&\dfrac{\mathcal{I}_{n}(\mu,\lambda,0)}{\log(2)}=\log_{2}(e)\exp\left(\dfrac{\mu}{\lambda}\right) E_{n}\left(\dfrac{\mu}{\lambda}\right).
\end{eqnarray}
We define a second integral  $\mathcal{J}(\mu,\lambda)$ as follows:
	\begin{eqnarray}\label{eqn:J_mu_lambda}
		\begin{array}{lll}\vspace{-0.1in}
			\mathcal{J}(\mu,\lambda)=  \displaystyle\int\limits_{0}^{\infty}\dfrac{\log_{2}(1+x)}{x+\mu}e^{-{x}/\lambda} {\D}x\\ \vspace{-0.1in}
			&& \hspace{-1.75in}=\exp\left(\dfrac{\mu}{\lambda}\right)\log_{2}(e)   \displaystyle\int\limits_{0}^{\infty}\dfrac{E_{1}\left(\dfrac{x+\mu}{\lambda}\right)}{1+x} {\D}x,
		\end{array}
	\end{eqnarray}
	where the last equality is obtained using integration by parts. $\mathcal{J}(\mu,\lambda)$ cannot be expressed in closed form. However, it can be approximated as follows:
		\begin{eqnarray*}
		\begin{array}{lll}
		\hspace{-0.3cm}\mathcal{J}(\mu,\lambda)\hspace{-0.2cm}&\approx&\hspace{-0.2cm}\dfrac{ \exp\left(\dfrac{\mu}{\lambda}\right)}{\log(2)}\Big[\dfrac{1}{2}\left\{EuM+\log\left(\dfrac{\mu}{\lambda}\right)\right\}^2+\dfrac{{\pi}^2}{12}\\
		&&\hspace{-1.8cm}+\
		\sum\limits_{k=1}^{\infty}\dfrac{\left(-\dfrac{\mu}{\lambda}\right)^k}{k^2k!}+\log(\mu) E_{1}\left(\dfrac{\mu}{\lambda}\right) \Big]-\log_{2}(e)\sum\limits_{k=1}^{\infty}\dfrac{\left(1-\dfrac{1}{\mu}\right)^k}{k(k+\dfrac{\mu}{\lambda})}.
		\end{array}
		\end{eqnarray*}
	where $EuM$ is the Euler-Mascheroni constant. Proof is omitted due to paucity of space. It can be shown that for $\lambda\rightarrow\infty$: 
	\begin{eqnarray*}
		\begin{array}{lll}
			\hspace{-0.3cm}\mathcal{J}(\mu,\lambda\rightarrow\infty)\hspace{-0.2cm}&\approx&\hspace{-0.2cm}\dfrac{1}{\log(2)}\Big[\dfrac{1}{2}\left(EuM-\log{\lambda}\right)^2+\dfrac{{\pi}^2}{12}+Di_{2}(\mu)\Big],
		\end{array}
	\end{eqnarray*}
	where $Di_{2}(x)$ is called Dilogarithm function\cite[27.7.2-5]{Abramowitz1965}.
	Now the achievable rate for ALSBR $\overline{{\mathcal{R}}}^{ALSBR}$ is evaluated using (\ref{eqn:rate_expr}). Unfortunately, an analytical expression for $\rho$ is not possible, and  numerical techniques are needed to evaluate $\rho$ that makes rates of the links equal.   Rate for SS-SR link $\mathbf{\E}_{\gamma_{s},\gamma_{r}}[(1-d)C_{s}]$ is given by:
		\vspace{-0.1in}	
	\begin{eqnarray*}
		\begin{array}{lll}\label{eqn:AER_EDNCS_GENERAL}
	\vspace{-0.10in}		E[(1-d)Cs]\hspace{-0.1in}&=&-\displaystyle\int\limits_{0}^{\infty} \log_{2}(1+x)\,{\D}F^c_{d,\gamma_{s}}(0,x)\\
			&&\hspace{-0.25in}=\
			\log_{2}(e)  \displaystyle\int\limits_{0}^{\infty} \dfrac{F^c_{d,\gamma_{s}}(0,x)}{1+x}\,{\D}x,
		\end{array}
	\end{eqnarray*}\vspace{-.35cm}
	\\where integration by parts is used to obtain the second equality. After substituting from (\ref{eqn:CCDF_SR}) and  using some manipulations, we get (\ref{eqn:ER_SR}). Similar expressions for rate  of SR-SD link i.e. $\mathbf{\E}_{\gamma_{s},\gamma_{r}}[d C_{r}]$ are obtained by exchanging variables as in (\ref{eqn:exchange}), which is given by (\ref{eqn:ER_RD}).
	\par The average rate of CUBR is given by (\ref{eqn:rate_expr_CUBR}). After averaging over end-to-end CCDF $F^c_{\gamma_{ete}}(x)=F^c_{\gamma_{s}}(x)F^c_{\gamma_{r}}(x)$,  $\overline{{\mathcal{R}}}^{CUBR}$ is given by (\ref{eqn:ER_CUBR}), where $\lambda_{e}$ is the end-to-end average SNR in non-cognitive scenario (PTPR)  given by the harmonic mean of $\lambda_{s}$ and $\lambda_{r}$ ($1/\lambda_{e}=1/\lambda_{s}+1/\lambda_{r}$). Note that $\lambda_{\rho}=\lambda_{e}$ for $\rho=1$. Proof is omitted due to space constraints. The average rate of CBR is given by (\ref{eqn:rate_expr_CBR}). Evaluating $\mathbf{\E}_{\gamma_{i}}[C_{i}]$ from (\ref{eqn:F_Gi_s}) and substituting, the average rate $\overline{{\mathcal{R}}}^{CBR}$ is given by (\ref{eqn:ER_CBR}). It can be verified that with $p_{s}=p_{r}=0$, the derived expressions reduce to the expressions for the cooeprative communications case presented in \cite{Zlatanov2013_1}.

	\begin{table*}[!t]
		%\caption{CCDF and Average-Rate}
		\caption {CCDF and Average-Rate} \label{tab:table_I} 
		\footnotesize
		\centering
		\renewcommand{\arraystretch}{1.5}
		\begin{minipage}{\textwidth}
			\begin{tabular}{|p{17.8cm}|}
				\hline\hline
				\vspace{-0.4cm}
				\begin{subequations}\label{eqn:CCDF_SR}
				\begin{eqnarray}
				\begin{array}{lll}
				\hspace{-0.05in}F_{d,\gamma_{s}}^c(0,x)=(1-p_{s})\Big[e^{-x/\lambda_{s}}-(1-p_{r})\dfrac{\lambda_{\rho}}{\rho \lambda_{s}}e^{-(\rho x)/\lambda_{\rho}}\Big]+ p_{s}\dfrac{\mu_{s}}{x+\mu_{s}}\Big[e^{-x/\lambda_{s}} - \Big(1-p_{r}+ \dfrac{\mu_{r}p_{r}}{\mu_{r}-\rho\mu_{s}}\Big)e^{-(\rho x)/\lambda_{\rho}}\Big]\\
				\vspace{0.0cm}
				\hspace{1.in}+p_{s} \left(1-p_{r}+  \dfrac{\mu_{r}p_{r}}{\mu_{r}-\rho\mu_{s}}+\dfrac{\lambda_{r}\,\mu_{r}p_{r}}{(\mu_{r}-\rho\mu_{s})^2}\right)\dfrac{\rho\mu_{s}}{\lambda_{r}} \exp\left(\dfrac{\rho \mu_{s}}{\lambda_{\rho}}\right)  E_{1}\left(\dfrac{\rho x+\rho \mu_{s}}{\lambda_{\rho}}\right)\\
				\vspace{0.0cm}
				\hspace{1.in}-p_{r}\left(1-p_{s}- \dfrac{\rho\mu_{s}p_{s}}{\mu_{r}-\rho\mu_{s}}+\dfrac{\rho\lambda_{s}\,\rho\mu_{s}\,p_{s}}{(\mu_{r}-\rho\mu_{s})^2}\right)\dfrac{\mu_{r}}{\rho\lambda_{s}} \exp\left(\dfrac{\mu_{r}}{\lambda_{\rho}}\right) E_{1}\left(\dfrac{\rho x+\mu_{r}}{\lambda_{\rho}}\right)\hspace{0.80in} \textbf{when}\,\,\bf{\mu_{r}\neq\rho\mu_{s}}.\label{eqn:CCDF_SR1}\\
				\end{array}
				\end{eqnarray}
				\vspace{-0.30cm}
				\hspace{-0.305cm}\text{---------------------------------------------------------------------------------------------------------------------------------------------------------------------------------------------------}\vspace{0.1cm}
				\begin{eqnarray}
				\begin{array}{lll}
				\hspace{-0.05in}F_{d,\gamma_{s}}^c(0,x) \hspace{-0. in}= (1-p_{s})\Big[e^{-x/\lambda_{s}}-(1-p_{r})\dfrac{\lambda_{\rho}}{\rho \lambda_{s}}e^{-(\rho x)/\lambda_{\rho}}\;\Big]-\dfrac{p_{s}p_{r}}{2}  e^{-(\rho x)/\lambda_{\rho}}\left(\dfrac{\mu_{s}}{x+\mu_{s}}\right)^2 \\
				\vspace{-0.cm}
				\hspace{1. in} +\
				\dfrac{p_{s}\mu_{s}}{x+\mu_{s}}\Big[e^{-x/\lambda_{s}}-(1-p_{r})e^{-(\rho x)/\lambda_{\rho}}+\dfrac{p_{r}}{2}\left(\dfrac{\mu_{r}}{\lambda_{r}}-\,\dfrac{\mu_{s}}{\lambda_{s}}\right)e^{-(\rho x)/\lambda_{\rho}} \Big]
				\\
				\vspace{-0.cm}
				\hspace{1. in} +\ \Big[p_{s}(1-p_{r})\dfrac{\mu_{r}}{\lambda_{r}}-p_{r}(1-p_{s})\dfrac{\mu_{s}}{\lambda_{s}}+\dfrac{p_{s}p_{r}}{2}\left(\dfrac{\mu_{s}^2}{\lambda_{s}^2}-\dfrac{\mu_{r}^2}{\lambda_{r}^2}\right)\Big]\exp\left(\dfrac{\mu_{r}}{\lambda_{\rho}}\right) E_{1}\left(\dfrac{\rho x+\mu_{r}}{\lambda_{\rho}}\right)\hspace{0.3in} \textbf{when}\,\,\bf{\mu_{r}=\rho\mu_{s}}.\label{eqn:CCDF_SR2}\\
				\end{array}
				\end{eqnarray}
				\end{subequations}
				\vspace{-0.25cm}\\\hline\vspace{-0.5cm}
				\begin{subequations}\label{eqn:ER_SR}
				\begin{eqnarray}
				\begin{array}{lll}
				\hspace{-0.05in}\mathbf{\E}_{\gamma_{s},\gamma_{r}}[(1-d)C_{s}]\hspace{-0.0in}  = 
				(1-p_{s}) \Big[\mathcal{I}_{1}(1,\lambda_{s})-(1-p_{r})\dfrac{\lambda_{\rho}}{\rho \lambda_{s}}\mathcal{I}_{1}\left(1,\dfrac{\lambda_{\rho}}{\rho}\right)\;\Big]+p_{s}\dfrac{\mu_{s}}{\mu_{s}-1}\\
				\vspace{-0.cm}
				\hspace{0.18in} \times\ 
				\Big[\mathcal{I}_{1}(1,\lambda_{s})-\mathcal{I}_{1}(\mu_{s},\lambda_{s})-\left(1-p_{r}+ \dfrac{\mu_{r}p_{r}}{\mu_{r}-\rho\mu_{s}}\right)\left\{\mathcal{I}_{1}\left(1,\dfrac{\lambda_{\rho}}{\rho}\right)-\mathcal{I}_{1}\left(\mu_{s},\dfrac{\lambda_{\rho}}{\rho}\right)\right\}\Big] +\dfrac{\rho\mu_{s}p_{s}}{\lambda_{r}}\mathcal{J}\left(\mu_{s},\dfrac{\lambda_{\rho}}{\rho}\right) \\
				\vspace{-0.cm}
				\hspace{0.18in} \times\ 
				\left(1-p_{r}+ \dfrac{\mu_{r}p_{r}}{\mu_{r}-\rho\mu_{s}}+\dfrac{\lambda_{r}\mu_{r}p_{r}}{(\mu_{r}-\rho\mu_{s})^2}\right) -\dfrac{\mu_{r}p_{r}}{\rho\lambda_{s}}\mathcal{J}\left(\dfrac{\mu_{r}}{\rho},\dfrac{\lambda_{\rho}}{\rho}\right)\left(1-p_{s}- \dfrac{\rho\mu_{s}p_{s}}{\mu_{r}-\rho\mu_{s}}+\dfrac{\rho\lambda_{s}\,\rho\mu_{s}\,p_{s}}{(\mu_{r}-\rho\mu_{s})^2}\right)\hspace{0.19in} \textbf{when}\,\,\bf{\mu_{r}\neq\rho\mu_{s}}.\label{eqn:ER_SR1}\\
				\end{array}
				\end{eqnarray}
				\vspace{-0.30cm}
				\hspace{-0.305cm}\text{---------------------------------------------------------------------------------------------------------------------------------------------------------------------------------------------------}\vspace{0.1cm}
				\begin{eqnarray}
				\begin{array}{lll}
				\mathbf{\E}_{\gamma_{s},\gamma_{r}}[(1-d)C_{s}]  =
				(1-p_{s}) \Big[\mathcal{I}_{1}(1,\lambda_{s})-(1-p_{r})\dfrac{\lambda_{\rho}}{\rho \lambda_{s}}\mathcal{I}_{1}\left(1,\dfrac{\lambda_{\rho}}{\rho}\right)\;\Big]+\dfrac{\mu_{s}p_{s}}{\mu_{s}-1}  \Big\{\mathcal{I}_{1}(1,\lambda_{s})-\mathcal{I}_{1}(\mu_{s},\lambda_{s})\Big\}\\
				\vspace{-0.cm}
				\hspace{.25in} -\ 
				\Big[\left\{p_{s}(1-p_{r})-\dfrac{p_{s}p_{r}}{2}\left(\dfrac{\mu_{r}}{\lambda_{r}}-\,\dfrac{\mu_{s}}{\lambda_{s}}\right)\right\}\dfrac{\mu_{s}}{\mu_{s}-1}+\dfrac{p_{s}p_{r}}{2}\left(\dfrac{\mu_{s}}{\mu_{s}-1}\right)^2\Big]\bigg\{\mathcal{I}_{1}\left(1,\dfrac{\lambda_{\rho}}{\rho}\right)-\mathcal{I}_{1}\left(\mu_{s},\dfrac{\lambda_{\rho}}{\rho}\right)\bigg\}\\
				\vspace{-0.cm}
				\hspace{.25in} +\ 
				\dfrac{p_{s}p_{r}}{2}\dfrac{\mu_{s}}{\mu_{s}-1}  \mathcal{I}_{2}\left(\mu_{s},\dfrac{\lambda_{\rho}}{\rho}\right)+\Big[p_{s}(1-p_{r})\dfrac{\mu_{r}}{\lambda_{r}}-p_{r}(1-p_{s})\dfrac{\mu_{s}}{\lambda_{s}}+\dfrac{p_{s}p_{r}}{2}\left(\dfrac{\mu_{s}^2}{\lambda_{s}^2}-\dfrac{\mu_{r}^2}{\lambda_{r}^2}\right)\Big] \mathcal{J}\Big(\mu_{s},\dfrac{\lambda_{\rho}}{\rho}\Big)\hspace{0.4in} \textbf{when}\,\,\bf{\mu_{r}=\rho\mu_{s}}.\label{eqn:ER_SR2}\\
				\end{array}
				\end{eqnarray}
				\end{subequations}
				\vspace{-0.25cm}\\\hline\vspace{-0.5cm}
				\begin{subequations}\label{eqn:ER_RD}
				\begin{eqnarray}
				\begin{array}{lll}
				\mathbf{\E}_{\gamma_{s},\gamma_{r}}[d C_{r}]  = 
				(1-p_{r}) \Big[\mathcal{I}_{1}(1,\lambda_{r})-(1-p_{s})\dfrac{\lambda_{\rho}}{\lambda_{r}}\mathcal{I}_{1}\left(1,\lambda_{\rho}\right)\;\Big]+p_{r}\dfrac{\mu_{r}}{\mu_{r}-1}\\
				\vspace{-0.cm}
				\hspace{.25in} \times\ 
				\Big[\mathcal{I}_{1}(1,\lambda_{r})-\mathcal{I}_{1}(\mu_{r},\lambda_{r})-\left(1-p_{s}- \dfrac{\rho\mu_{s}p_{s}}{\mu_{r}-\rho\mu_{s}}\right)\left\{\mathcal{I}_{1}\left(1,\lambda_{\rho}\right)-\mathcal{I}_{1}\left(\mu_{r},\lambda_{\rho}\right)\right\}\Big] -\dfrac{\rho\mu_{s}p_{s}}{\lambda_{r}}\mathcal{J}\left(\mu_{r},\lambda_{\rho}\right) \\
				\vspace{-0.cm}
				\hspace{0.25in} \times\ 
				\left(1-p_{r}+ \dfrac{\mu_{r}p_{r}}{\mu_{r}-\rho\mu_{s}}+\dfrac{\lambda_{r}\,\mu_{r}\,p_{r}}{(\mu_{r}-\rho\mu_{s})^2}\right) -\dfrac{\mu_{r}p_{r}}{\rho\lambda_{s}}\mathcal{J}(\rho\mu_{s},\lambda_{\rho})\left(1-p_{s}- \dfrac{\rho\mu_{s}p_{s}}{\mu_{r}-\rho\mu_{s}}+\dfrac{\rho\lambda_{s}\,\rho\mu_{s}\,p_{s}}{(\mu_{r}-\rho\mu_{s})^2}\right)\hspace{0.22in} \textbf{when}\,\,\bf{\mu_{r}\neq\rho\mu_{s}}.\label{eqn:ER_RD1}\\
				\end{array}
				\end{eqnarray}
				\vspace{-0.30cm}
				\hspace{-0.305cm}\text{---------------------------------------------------------------------------------------------------------------------------------------------------------------------------------------------------}\vspace{0.1cm}
				\begin{eqnarray}
				\begin{array}{lll}
				\mathbf{\E}_{\gamma_{s},\gamma_{r}}[d C_{r}]  =
				(1-p_{r}) \Big[\mathcal{I}_{1}(1,\lambda_{r})-(1-p_{s})\dfrac{\lambda_{\rho}}{ \lambda_{r}}\mathcal{I}_{1}(1,\lambda_{\rho})\;\Big]+\dfrac{\mu_{r}p_{r}}{\mu_{r}-1} \Big\{\mathcal{I}_{1}(1,\lambda_{r})-\mathcal{I}_{1}(\mu_{r},\lambda_{r})\Big\}\\
				\vspace{-0.cm}
				\hspace{.25in} -\ 
				\Big[\left\{p_{r}(1-p_{s})-\dfrac{p_{s}p_{r}}{2}\left(\dfrac{\mu_{s}}{\lambda_{s}}-\,\dfrac{\mu_{r}}{\lambda_{r}}\right)\right\}\dfrac{\mu_{r}}{\mu_{r}-1}+\dfrac{p_{s}p_{r}}{2}\left(\dfrac{\mu_{r}}{\mu_{r}-1}\right)^2\Big]\Big\{\mathcal{I}_{1}(1,\lambda_{\rho})-\mathcal{I}_{1}(\mu_{r},\lambda_{\rho})\Big\}\\
				\vspace{-0.cm}
				\hspace{.25in} +\ 
				\dfrac{p_{s}p_{r}}{2}\dfrac{\mu_{r}}{\mu_{r}-1}  \mathcal{I}_{2}\left(\mu_{r},\lambda_{\rho}\right)+\Big[p_{r}(1-p_{s})\dfrac{\mu_{s}}{\lambda_{s}}-p_{s}(1-p_{r})\dfrac{\mu_{r}}{\lambda_{r}}+\dfrac{p_{s}p_{r}}{2}\left(\dfrac{\mu_{r}^2}{\lambda_{r}^2}-\dfrac{\mu_{s}^2}{\lambda_{s}^2}\right)\Big] \mathcal{J}(\mu_{r},\lambda_{\rho})\hspace{.5in} \textbf{when}\,\,\bf{\mu_{r}=\rho\mu_{s}}.\label{eqn:ER_RD2}\\
				\end{array}
				\end{eqnarray}
				\end{subequations}
				\vspace{-0.25cm}\\\hline\vspace{-0.5cm}
				\begin{subequations}\label{eqn:ER_CUBR}
				\begin{eqnarray}
				\begin{array}{lll}
				\hspace{-0.1in}\overline{{\mathcal{R}}}^{CUBR}\hspace{-0.15in}  &=& 1/2   \Big[(1-p_{s})(1-p_{r})\mathcal{I}_{1}(1,\lambda_{e})+\dfrac{p_{s}\mu_{s}}{\mu_{s}-1}\left(1-p_{r}+p_{r}\dfrac{\mu_{r}}{\mu_{r}-\mu_{s}}\right)\Big\{ \mathcal{I}_{1}(1,\lambda_{e})-\mathcal{I}_{1}(\mu_{s},\lambda_{e})\Big\}  \\
				&& \hspace{.8in} +\ \dfrac{p_{r}\mu_{r}}{\mu_{r}-1}\left(1-p_{s}-p_{s}\dfrac{\mu_{s}}{\mu_{r}-\mu_{s}}\right)\Big\{ \mathcal{I}_{1}(1,\lambda_{e})-\mathcal{I}_{1}(\mu_{r},\lambda_{e})\Big\}\Big]\hspace{1.in} \textbf{when}\,\,\bf{\mu_{r}\neq\rho\mu_{s}}.\label{eqn:ER_CUBR1}\\
				\end{array}
				\end{eqnarray}
				\vspace{-0.30cm}
				\hspace{-0.305cm}\text{---------------------------------------------------------------------------------------------------------------------------------------------------------------------------------------------------}\vspace{0.1cm}
				\begin{eqnarray}
				\begin{array}{lll}
				\hspace{-0.1in}\overline{{\mathcal{R}}}^{CUBR}\hspace{-0.15in}  &=&\hspace{-0.1in} 1/2  \Big[ (1-p_{s})(1-p_{r})\mathcal{I}_{1}(1,\lambda_{e})-p_{s}p_{r}\dfrac{\mu_{s}}{\mu_{s}-1} \mathcal{I}_{2}\left(\mu_{s},\lambda_{e}\right)+  \Big[\Big\{p_{s}(1-p_{r})+p_{r}(1-p_{s})\Big\}\dfrac{\mu_{s}}{\mu_{s}-1}+ p_{s}p_{r}\left(\dfrac{\mu_{s}}{\mu_{s}-1}\right)^2\Big]\\
				&& \hspace{0.7in} \times\
				\Big\{ \mathcal{I}_{1}(1,\lambda_{e})-\mathcal{I}_{1}(\mu_{s},\lambda_{e})\Big\}\Big].\hspace{2.68in} \textbf{when}\,\,\bf{\mu_{r}=\rho\mu_{s}}.\label{eqn:ER_CUBR2}\\
				\end{array}
				\end{eqnarray}
				\end{subequations}
				
				\vspace{-0.35cm}\\\hline\vspace{-0.4cm}
				\begin{eqnarray}
				\begin{array}{lll}
				\hspace{.25cm}\overline{{\mathcal{R}}}^{CBR}\hspace{-0.15in}  &=&\hspace{-0.1in} 1/2  \min \Big[(1-p_{s}) \mathcal{I}_{1}(1,\lambda_{s})+\dfrac{p_{s}\mu_{s}}{\mu_{s}-1}
				\Big\{\mathcal{I}_{1}(1,\lambda_{s})-\mathcal{I}_{1}(\mu_{s},\lambda_{s})\Big\},(1-p_{r}) \mathcal{I}_{1}(1,\lambda_{r})+\dfrac{p_{r}\mu_{r}}{\mu_{r}-1}
				\Big\{\mathcal{I}_{1}(1,\lambda_{r})-\mathcal{I}_{1}(\mu_{r},\lambda_{r})\Big\}\Big].\label{eqn:ER_CBR}
				\end{array}
				\end{eqnarray}
				
				\vspace{-0.375cm}\\\hline\hline
			\end{tabular}
			\label{tab:CCDF_ER} 
		\end{minipage}
	\end{table*}	
	\subsubsection*{\textbf{High SNR Average Rate}}
	\par We now derive approximate expressions for the average rate of ALSBR at high SNRs ($p_{s}=p_{r}=1$), that corresponds to the PIPR case (which implies large $\lambda_{s}$ and $\lambda_r$). In (\ref{eqn:I_n_mu_lambda}), using 
	 $\E_{1}(x)\approx\log(1/x)$ for small $x$, we get from (\ref{eqn:I_n_mu_lambda}) $\mathcal{I}_{1}(\mu,\lambda)\approx\log_{2}(\lambda/\mu)$. Applying this approximation in (\ref{eqn:ER_SR1}), it can be seen that the average rate of SS-SR link in PIPR can be written as:
	\begin{eqnarray}
		\begin{array}{lll}\nonumber
			\hspace{-.6cm}\mathbf{\E}_{\gamma_{s},\gamma_{r}}^{asym}[(1-d)C_{s}]\hspace{-0.1in}&=&\hspace{-0.1in} \dfrac{-\rho\mu_{s}}{\mu_{r}-\rho\mu_{s}}\dfrac{\mu_{s}}{\mu_{s}-1}\log_{2}(\mu_{s})\\
			&&\hspace{-2.15cm} +\
			\dfrac{\rho\mu_{s}\mu_{r}}{(\mu_{r}-\rho\mu_{s})^2} \Big[\mathcal{{J}}(\mu_{s},\infty)-\mathcal{{J}}(\mu_{r}/\rho,\infty)\Big],
		\end{array}
	\end{eqnarray}
	\vspace{-0.5cm}
	\begin{eqnarray}\label{eqn:AER_ENDCS1_PIPR}
	\begin{array}{lll}
	\hspace{-.1cm}\mathbf{\E}_{\gamma_{s},\gamma_{r}}^{asym}[(1-d)C_{s}]\hspace{-0.1in}&=&
	\dfrac{-\rho\mu_{s}}{\mu_{r}-\rho\mu_{s}}\dfrac{\mu_{s}}{\mu_{s}-1}\log_{2}(\mu_{s})\\
	&&\hspace{-2.15cm} +\
	\dfrac{\rho\mu_{s}\mu_{r}}{(\mu_{r}-\rho\mu_{s})^2}\log_{2}(e)\Big\{Di_{2}(\mu_{s})-Di_{2}(\dfrac{\mu_{r}}{\rho})\Big\},
	\end{array}	\vspace{-0.1in}
	\end{eqnarray}
where last line can be obtained by using the approximation of (\ref{eqn:J_mu_lambda}). For the special case when $\mu_{r}=\rho\mu_{s}$, we can apply similar approximations starting with (\ref{eqn:ER_SR2}) to get:
	\begin{eqnarray}\label{eqn:AER_ENDCS2_PIPR}
	%\nonumber
	\hspace{-.23cm}\mathbf{\E}_{\gamma_{s},\gamma_{r}}^{asym}[(1-d)C_{s}]\hspace{-0.13in}&=&\hspace{-0.14in} \dfrac{0.5\mu_{s}}{\mu_{s}-1}\Big[\hspace{-0.05in}\log_{2}(e)\hspace{-0.05in}+\hspace{-.1cm}\left(\hspace{-.05cm}\frac{\mu_{s}-2}{\mu_{s}-1}\hspace{-.05cm}\right)\hspace{-0.05in}\log_2(\mu_{s})\Big]
	\vspace{-0.2in}
	\end{eqnarray}
	$\mathbf{\E}_{\gamma_{s},\gamma_{r}}^{asym}[dC_{r}]$ can be obtained from (\ref{eqn:AER_ENDCS2_PIPR}) using (\ref{eqn:exchange}).
	Using (\ref{eqn:I_n_mu_lambda}) and following a similar procedure,  the asymptotic average rate of CUBR in PIPR  when $\mu_r\neq \mu_s$ can be shown using (\ref{eqn:ER_CUBR1}) to be:
	\begin{eqnarray}
	\label{eqn:AER_CONV1_PIPR}
	\begin{array}{lll}
	\hspace{-0.13in}\overline{{\mathcal{R}}}^{CUBR}_{asym}\hspace{-0.15in}  &=&\hspace{-0.1in} \dfrac{1}{2} \left(\dfrac{\mu_{s}\mu_{r}}{\mu_{r}-\mu_{s}}\right)  \Big[\dfrac{\mu_{s}\log_{2}(\mu_{s})}{\mu_{s}-1}-\dfrac{\mu_{r}\log_{2}(\mu_{r})}{\mu_{r}-1} \Big].
	\end{array}
	\end{eqnarray}
When $\mu_{r}=\mu_{s}$, $\overline{{\mathcal{R}}}^{CUBR}_{asym}$ can be shown using (\ref{eqn:ER_CUBR2}) to be:
	\begin{eqnarray}
	\label{eqn:AER_CONV2_PIPR}
	\begin{array}{lll}\hspace{-0.1in}
	\overline{{\mathcal{R}}}^{CUBR}_{asym}\hspace{-0.15in}  &=&\hspace{-0.125in} \dfrac{1}{2} \log_{2}(e) \Big[ -\dfrac{\mu_{s}}{\mu_{s}-1} +\left(\dfrac{\mu_{s}}{\mu_{s}-1}\right)^2 \log(\mu_{s})\Big].
	\end{array}
	\end{eqnarray}
	$\overline{{\mathcal{R}}}^{CBR}_{asym}$ in PIPR can be found from (\ref{eqn:ER_CBR}) as:
	\begin{eqnarray}\label{eqn:ER_CBR_PIPR}
	\begin{array}{lll}
	\label{eqn:AER_Ei}
	\hspace{-0.1in}\overline{{\mathcal{R}}}^{CBR}_{asym}\hspace{-0.1in}  &=& \hspace{-0.1in}  \dfrac{1}{2} \min\Big\{\dfrac{\mu_{s}}{\mu_{s}-1}	\log_{2}(\mu_{s}),\dfrac{\mu_{r}}{\mu_{r}-1}	\log_{2}(\mu_{r})\Big\}.
	\end{array}
	\end{eqnarray}
When $\mu_{s}=\mu_{r}$, average rates of SS-SR \& SR-SD are the same. 
%	\vspace{-0.5cm}
%	\begin{eqnarray}
%	\begin{array}{lll}
%	\label{eqn:AER_Ei}
%	\hspace{-0.1in}\text{when}\, \mu_{r}=\mu_{s}\,\, \overline{{\mathcal{R}}}^{CBR}_{asym}= \dfrac{1}{2}\dfrac{\mu_{s}}{\mu_{s}-1}	\log_{2}(\mu_{s}). 
%	\end{array}
%	\end{eqnarray}
	
	\vspace{-0.5cm}\section{Simulation Results}\label{sec:SimRes}
	\begin{figure}[t]
		\begin{center}
			\includegraphics[scale=0.5]{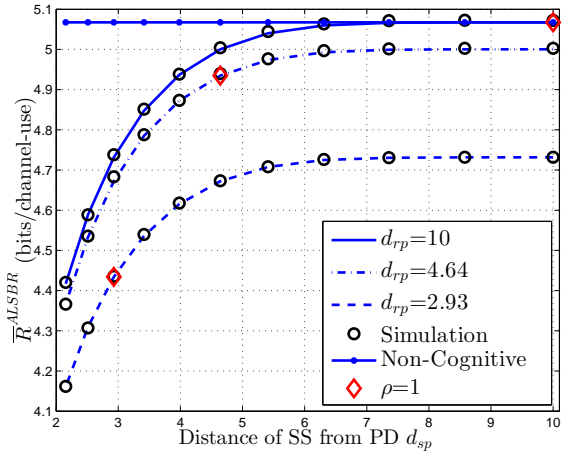}
			\caption{Achievable Rate of ALSBR vs $d_{sp}$, cf. (\ref{eqn:rate_expr})
				$\gamma_{max}=30\,dB$, $\gamma_{p}=10\,dB\,,\,\Omega_{hs}=\Omega_{hr}=1$.}
			\label{fig:OptimumThroughput}
		\end{center}
		\vspace{-.5cm}
	\end{figure}
	\begin{figure}[t]
		\begin{center}
			\includegraphics[scale=0.5]{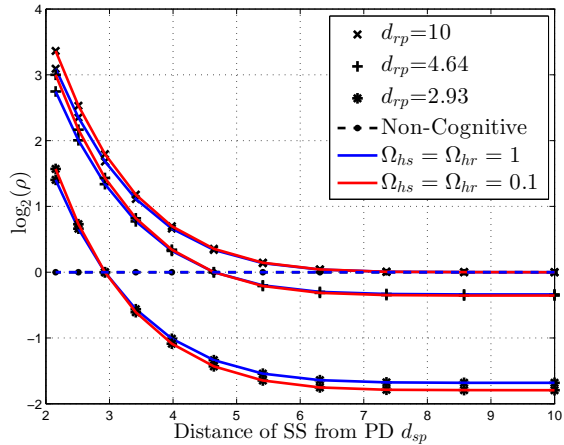}
			\caption{Link Selection Parameter $\rho$ of ALSBR vs $d_{sp}$, cf. (\ref{eqn:rate_expr}) $\gamma_{max}=30\, dB$, $\gamma_{p}=10\,dB$, $\alpha=3$.}
			\label{fig:OptimumRho} 
		\end{center}
		\vspace{-.5cm}
	\end{figure}
%	\vspace{-.3cm}
	In this section,  we present computer simulations to validate the presented analysis.  We assume $\gamma_{max}=30\,dB$, $\gamma_{p}=10\,dB$, and pathloss exponent $\alpha=3$. $d_{sp}$ is varied with $d_{rp}=10$ $d_{rp}=4.64$, and $d_{rp}=2.93$. %These correspond to 
	%$d_{sp}$ is selected such that $\mu_{s}/\lambda_{s}$ varies 10 to -10 $dB$ as it decreases whereas $\mu_{r}/\lambda_{r}$ is 10, 0 and -6 $dB$ for respective $d_{rp}$. 
	%$\mu_{r}=40, 30, 24\,dB$ for $\Omega_{hs}=\Omega_{hr}=1$, and  $\mu_{r}=30, 20, 14\,dB$ for $\Omega_{hs}=\Omega_{hr}=0.1$.
	%For Fig.~\ref{fig:OptimumThroughput}- Fig.~\ref{fig:OptimumRho}, we take SS, SR and SD at unit distance ($\Omega_{h_{s}}=\Omega_{h_{r}}=1$). For degrading SS-SR and SR-SD link, we take $\Omega_{h_{s}}=\Omega_{h_{r}}=0.1$ with pathloss coefficient $\alpha=3$. 
	\begin{figure}[t]
		\begin{center}
			\includegraphics[scale=0.5]{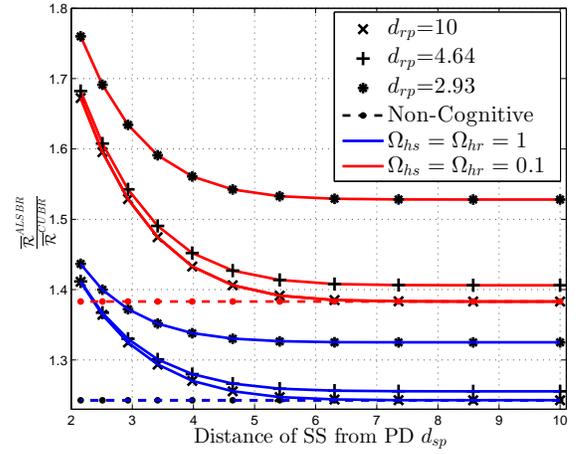}
			\caption{ Ratio of Achievable Rate of ALSBR w r t CUBR, cf. (\ref{eqn:rate_expr}), (\ref{eqn:ER_CUBR1}), (\ref{eqn:ER_CUBR2}), $\gamma_{max}=30\,dB$, $\gamma_{p}=10\,dB$, $\alpha=3$.} 
			\label{fig:IncreaseInThroughput}
		\end{center}
		\vspace{-.5cm}
	\end{figure}
	\begin{figure}[t]
		\begin{center}
			\includegraphics[scale=0.5]{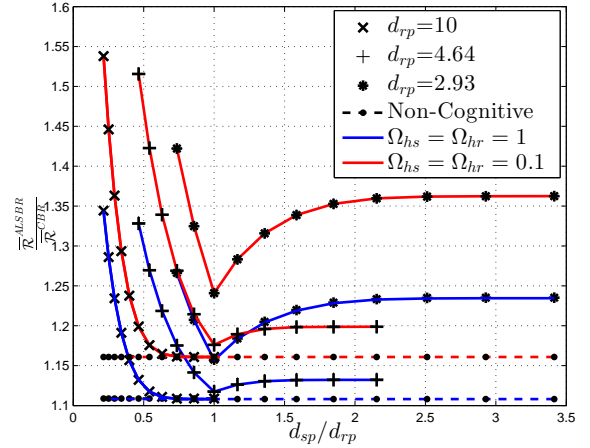}
			\caption{ Ratio of Achievable Rate of ALSBR w r t CBR, cf. (\ref{eqn:rate_expr}), (\ref{eqn:ER_CBR}) $\gamma_{max}=30\,dB$, $\gamma_{p}=10\,dB$, $\alpha=3$.} 
			\label{fig:IncreaseInThroughput3}
		\end{center}
		\vspace{-.5cm}
	\end{figure}
	 \par Fig.~\ref{fig:OptimumThroughput} depicts the average rate of the ALSBR versus $d_{sp}$ (the distance of SS from PD) for various $d_{rp}$ values, cf. (\ref{eqn:rate_expr}) where $\mathbf{\E}_{\gamma_{s},\gamma_{r}}[(1-d)C_{s}]$ is obtained from (\ref{eqn:ER_SR1}) or (\ref{eqn:ER_SR2}) and $\mathbf{\E}_{\gamma_{s},\gamma_{r}}[d C_{r}]$ is obtained from (\ref{eqn:ER_RD1}) or (\ref{eqn:ER_RD2}). It can be seen that for the same $d_{rp}$, the average rate saturates for higher $d_{sp}$ and does not improve further unless $d_{rp}$ is increased (thereby improving second hop performance). When both $d_{rp}$ and $d_{sp}$ are large, the system model becomes close to the non-cognitive scenario \cite{Zlatanov2013_1} as shown.
	 \par In Fig.~\ref{fig:OptimumRho}, the optimum value of $\log_{2}(\rho)$ chosen to satisfy (\ref{eqn:rate_expr}) is plotted versus $d_{sp}$ for various $d_{rp}$. When $d_{rp}>d_{sp}$ ($d_{rp}<d_{sp}$), $\rho>1$ ($\rho<1$).  As $d_{rp}$ decreases (so that SR-SD is the bottleneck link), $\rho$ decreases too, which demonstrates that  SS-SR link is selected  less frequently to ensure buffer stability. On the contrary, $\rho$ increases when $d_{sp}$ decreases. When $d_{sp}=d_{rp}$,   $\rho=1$.
	 \par Fig.~\ref{fig:IncreaseInThroughput} depicts the rate improvement of ALSBR  wrt CUBR (cf. (\ref{eqn:rate_expr}), (\ref{eqn:ER_CUBR1}) and (\ref{eqn:ER_CUBR2})), which demonstrates that the average rate ratio monotonically increases with the stronger of interference constraints. 
	 \par In Fig.~\ref{fig:IncreaseInThroughput3}, the ratio of rates  of ALSBR to that of CBR is plotted versus $d_{sp}/d_{rp}$, cf. (\ref{eqn:rate_expr}) and (\ref{eqn:ER_CBR}). It is clear that the ratio saturates for larger $d_{sp}$ and has a minimum when $d_{sp}=d_{rp}$. Although the average rate itself decreases, the ratio always improves when the channel between SS-SR and SR-SD degrades for both CUBR and CBR. 
	 
%	   The following important inference can be made: adaptive link selection is important to improve performance in interference dominated environments. 

%	 In both the cases, average rate always decreases due to interference constraints. Although the  average rate decreases, it is evident from Fig.~\ref{fig:IncreaseInThroughput} that the ratio of average rates of ALSBR and CUBR increases with the stronger interference constraints. It is clear that there is larger improvement when  one link is in PIPR regime and the other in PTPR, as compared to the case when both are in PIPR regime.. These gains further increase when channel variances of SS-SR and/or SR-RD link decrease.

%	\par In ALSBR, the link having the maximum SNR is picked. For this reason, the  average rate is upper bounded by half of expectation of maximum of instantaneous capacities so that: $\overline{{\mathcal{R}}}^{ALSBR} \leq 1/2\,\mathbf{\E}_{\gamma_{s},\gamma_{r}}[\max\{C_{s},C_{r}\}]$. Equality occurs for identical CCDF of instantaneous SNR of both the link (which makes  $\rho=1$).
	
	\vspace{-0.1cm}\section{Conclusion}\label{sec:ConClu}	\vspace{-0.1cm}
	In this paper, rate performance of cognitive two-hop network using a buffer-aided decode and forward relay that uses adaptive link selection is analyzed. It is shown that adaptive link selection is of atmost importance in interference constrained underlay cognitive radio scenarios. This insight is useful to system designers. We derived expressions for average rate of the adaptive link selection scheme and compared the same  with conventional buffered and unbuffered schemes. 
	%All the analysis are for adaptive rate transmission, where infinite-sized buffer is used. It is possible to use finite-sized starving buffer for analysis and fixed-rate performance as discussed in the longer version of this paper.
	\section*{Acknowledgement}
	This work was supported by Information Technology Re-search Academy (a unit of Media Labs Asia) through spon- sored project ITRA/15(63)/Mobile/MBSSCRN/01.
	\section*{Appendix A}\vspace{-0.15cm}
	The derivation of (\ref{eqn:CCDF_SR1}) and (\ref{eqn:CCDF_SR2}) is presented in this Appendix. We use integral (\ref{eqn:I_n_mu_lambda_x}) in the derivation extensively. It can be shown that the integral obeys the following recursion relation:\vspace{-0.25cm}
	\begin{eqnarray*}
		\begin{array}{lll}
		\mathcal{I}_{n}(\mu,\lambda;x)\hspace{-0.025in}= \dfrac{1}{n-1}\left[e^{-{x}/{\lambda}}\left(\dfrac{\mu}{x+\mu}\right)^{n-1}\hspace{-0.15in}-\dfrac{\mu}{\lambda}\mathcal{I}_{n-1}(\mu,\lambda;x)\right].
		\end{array}{}
	\end{eqnarray*}\\\vspace{0.2cm}
	We know that
	$\hspace{-0.0in}F^c_{d,\gamma_{s}}(0,x)= F^c_{\gamma_{s}}(x)-F^c_{d,\gamma_{s}}(1,x)$ where CCDF $F^c_{\gamma_{s}}(x)$ is given by (\ref{eqn:F_Gi_s}). Now
	\small{\vspace{-.25cm}
		\begin{eqnarray*}
			\begin{array}{lll}\vspace{-0.2cm}
				F^c_{d,\gamma_{s}}(1,x)=\Pr\{\frac{\gamma_r}{\rho}>\gamma_s>x\}=\displaystyle\int\limits_{x}^{\infty}F^c_{\gamma_{r}}(\rho s)f_{\gamma_{s}}(s) {\D}s.
			\end{array}
		\end{eqnarray*}
	}
	\normalsize{Substituting from (\ref{eqn:F_Gi_s}) and (\ref{eqn:f_Gi_s}), we get:}
	\small{
	\begin{eqnarray*}
		\begin{array}{lll}
			\hspace{-0.1 in}F^c_{d,\gamma_{s}}(1,x)\hspace{-0.1 in}&=&  \dfrac{1}{\lambda_{s}}\displaystyle\int\limits_{x}^{\infty} e^{-(\rho s)/\lambda_{r}}\left[1-p_{r}\left(1- \dfrac{\mu_{r}}{\rho s+\mu_{r}}\right) \right]\\
			&&\hspace{-0.8 in} \times\
			e^{-s/\lambda_{s}}\left[1-p_{s}\left(1- \dfrac{\mu_{s}}{s+\mu_{s}}- \dfrac{\lambda_{s}\mu_{s}}{(s+\mu_{s})^2}\right) \right]{\D}s\\
			&&\hspace{-0.8 in} =\
			 \underbrace{ \dfrac{1}{\lambda_{s}}\int\limits_{x}^{\infty} e^{-(\rho\,s)/\lambda_{\rho}} {\D}s}_{T_{1}}-p_{r}   \underbrace{\dfrac{1}{\lambda_{s}}\int\limits_{x}^{\infty} e^{-(\rho\,s)/\lambda_{\rho}}\left(1- \dfrac{\mu_{r}}{\rho s+\mu_{r}}\right) {\D}s}_{T_{4}}\\
			&&\hspace{-0.8 in} -\
			p_{s} \underbrace{\dfrac{1}{\lambda_{s}}\int\limits_{x}^{\infty} e^{-(\rho\,s)/\lambda_{\rho}}\left(1- \dfrac{\mu_{s}}{s+\mu_{s}}- \dfrac{\lambda_{s}\mu_{s}}{(s+\mu_{s})^2}\right) {\D}s}_{T_{2}-T_{3}}+ \,p_{s}p_{r} \\
			&&\hspace{-0.8 in} \times\
			 \underbrace{\dfrac{1}{\lambda_{s}}\int\limits_{x}^{\infty} e^{-(\rho\,s)/\lambda_{\rho}}\left[1- \dfrac{\mu_{s}}{s+\mu_{s}}- \dfrac{\lambda_{s}\mu_{s}}{(s+\mu_{s})^2}\right]\left[1- \dfrac{\mu_{r}}{\rho s+\mu_{r}}\right] {\D}s}_{(T_{2}-T_{3})-T_{5}} \\
			&&\hspace{-0.8 in} =\
			T_{1} - p_{s} (T_{2}-T_{3})-p_{r} T_{4} + p_{s}p_{r}\{(T_{2}-T_{3})-T_{5}\}
		\end{array}
	\end{eqnarray*}
	}
	\normalsize{where the last line is obtained by collecting $p_{s}$, $p_{r}$ and $p_{s}p_{r}$ terms together. We now present expressions for each of the integrals $T_1$ - $T_5$. It can be shown that $T_1$ - $T_4$ are given by:}
	\small{
\begin{eqnarray*}
	\begin{array}{lll}
		T_{1}=\dfrac{1}{\lambda_{s}}\displaystyle\int\limits_{x}^{\infty} e^{-(\rho\,s)/\lambda_{\rho}} {\D}s =\dfrac{\lambda_{\rho}}{\rho\lambda_{s}} e^{-(\rho\,x)/\lambda_{\rho}}\\
		&&\hspace{-2.525in} T_{2}=\dfrac{\rho}{\lambda_{\rho}}\hspace{-0.1cm}\displaystyle\int\limits_{x}^{\infty}\hspace{-0.05cm} \left(1- \dfrac{\mu_{s}}{s+\mu_{s}}- \dfrac{\lambda_{\rho}\mu_{s}}{\rho(s+\mu_{s})^2}\right) e^{-(\rho\,s)/\lambda_{\rho}} {\D}s\overset{l}{=} \dfrac{x\,e^{-(\rho\,x)/\lambda_{\rho}}}{x+\mu_{s}}\\
		&&\hspace{-2.525in}T_{3}=\dfrac{\rho}{\lambda_{r}}\displaystyle\int\limits_{x}^{\infty} \left(1-\dfrac{\mu_{s}}{s+\mu_{s}}\right)  e^{-(\rho\,s)/\lambda_{\rho}} {\D}s\\
		&&\hspace{-2.35in} \overset{m}{=} \dfrac{\lambda_{\rho}}{\lambda_{r}}\left[e^{-(\rho\,x)/\lambda_{\rho}}-\dfrac{\rho\mu_{s}}{\lambda_{\rho}} \exp\left(\dfrac{\rho\mu_{s}}{\lambda_{\rho}}\right) E_{1}\left(\dfrac{\rho x+\rho\mu_{s}}{\lambda_{\rho}}\right)\right]\\
		&&\hspace{-2.525in}T_{4}=\dfrac{1}{\lambda_{s}}\displaystyle\int\limits_{x}^{\infty} \left(1-\dfrac{\mu_{r}}{\rho s+\mu_{r}}\right)  e^{-(\rho\,s)/\lambda_{\rho}} {\D}s\\
		&&\hspace{-2.35in} \overset{n}{=} \dfrac{\lambda_{\rho}}{\rho\lambda_{s}}\left[e^{-(\rho\,x)/\lambda_{\rho}}-\dfrac{\mu_{r}}{\lambda_{\rho}}\exp\left(\dfrac{\mu_{r}}{\lambda_{\rho}}\right)E_{1}\left(\dfrac{\rho x+\mu_{r}}{\lambda_{\rho}}\right)\right]. %\\
		%		&&\hspace{-2.52in}T_{5}\overset{p}{=}(T_{2}-T_{3})-T_{5},
	\end{array}
\end{eqnarray*}
}
\normalsize{Equality $l$ is derived using (\ref{eqn:I_n_mu_lambda_x}) and its recursion whereas equality $m$ and $n$ use only (\ref{eqn:I_n_mu_lambda_x}). Generally, $T_{5}$ is given as:}
\small{
\begin{eqnarray*}
	\begin{array}{lll}
		\hspace{-.7in}T_{5}=\dfrac{1}{\lambda_{s}}\displaystyle\int\limits_{x}^{\infty} e^{-(\rho\,s)/\lambda_{\rho}}\left(1- \dfrac{\mu_{s}}{s+\mu_{s}}- \dfrac{\lambda_{s}\mu_{s}}{(s+\mu_{s})^2}\right)\dfrac{\mu_{r}}{\rho s+\mu_{r}} {\D}s\\
		&& \hspace{-3.35in}\overset{p}{=}\
		\dfrac{\mu_{r}}{\mu_{r}-\rho\mu_{s}}\left[(T_{2}-T_{3})-T_{4}-T_{6}\right], \text{where}\vspace{-0.3cm}
	\end{array}
\end{eqnarray*}
\begin{eqnarray*}
	\begin{array}{lll}\vspace{-.3cm}		
		\hspace{-.4in}T_{6}=\displaystyle\int\limits_{x}^{\infty}\dfrac{\rho\mu_{s}}{(s+\mu_{s})(\rho s+\mu_{r})} e^{-(\rho\,s)/\lambda_{\rho}} {\D}s\overset{q}{=}\dfrac{\rho\mu_{s}}{\mu_{r}-\rho\mu_{s}}\\
	\end{array}
\end{eqnarray*}
\begin{eqnarray*}
	\begin{array}{lll}
		&&\hspace{-.3in} \times\
		\left[\exp\left(\dfrac{\rho\mu_{s} }{\lambda_{\rho}}\right)E_{1}\left(\dfrac{\rho x+\rho\mu_{s}}{\lambda_{\rho}}\right)-\exp\left(\dfrac{\mu_{r} }{\lambda_{\rho}}\right)E_{1}\left(\dfrac{\rho x+\mu_{r}}{\lambda_{\rho}}\right)\right],
	\end{array}
\end{eqnarray*}
}
\normalsize{In the above, equality $p$ and $q$ result from partial fraction expansion and some manipulation using (\ref{eqn:I_n_mu_lambda_x}).Under particular condition when $\mu_{r}=\rho\mu_{s}$, $T_{5}$ is given as:}\vspace{-0.25cm}
\small{
\begin{eqnarray*}
	\begin{array}{lll}
		T_{5}\hspace{-.02in}=\dfrac{1}{\lambda_{s}}\displaystyle\int\limits_{x}^{\infty} e^{-(\rho\,s)/\lambda_{\rho}}\left(1- \dfrac{\mu_{s}}{s+\mu_{s}}- \dfrac{\lambda_{s}\mu_{s}}{(s+\mu_{s})^2}\right)\dfrac{\mu_{s}}{s+\mu_{s}} {\D}s\\
		&&\hspace{-3.31in}\overset{r}{=}\
		\dfrac{\mu_{s}}{\lambda_{s}}\exp\left(\dfrac{\mu_{r}}{\lambda_{\rho}}\right)E_{1}\left(\dfrac{\rho x+\mu_{r}}{\lambda_{\rho}}\right)-\dfrac{1}{2}e^{-(\rho\,x)/\lambda_{\rho}}\bigg[\left(\dfrac{\mu_{s}}{x+\mu_{s}}\right)^2\\
		&&\hspace{-3.31in}
		-\left(\dfrac{\mu_{r}}{\lambda_{r}}-\dfrac{\mu_{s}}{\lambda_{s}}\right)\left[\dfrac{\mu_{s}}{x+\mu_{s}}-\dfrac{\rho\mu_{s}}{\lambda_{\rho}}\exp\left(\dfrac{\mu_{r}}{\lambda_{\rho}}\right)E_{1}\left(\dfrac{\rho x+\mu_{r}}{\lambda_{\rho}}\right)\right]\bigg]
	\end{array}
\end{eqnarray*}
}
\normalsize{
Equality $r$ is established using (\ref{eqn:I_n_mu_lambda_x}) after some manipulation.
After rearranging all the terms, we get  (\ref{eqn:CCDF_SR}).
}\\
	\bibliographystyle{ieeetr}
	\bibliography{ARXIV_Sarnoff}

\begin{thebibliography}{10}
\providecommand{\url}[1]{#1}
\csname url@samestyle\endcsname
\providecommand{\newblock}{\relax}
\providecommand{\bibinfo}[2]{#2}
\providecommand{\BIBentrySTDinterwordspacing}{\spaceskip=0pt\relax}
\providecommand{\BIBentryALTinterwordstretchfactor}{4}
\providecommand{\BIBentryALTinterwordspacing}{\spaceskip=\fontdimen2\font plus
\BIBentryALTinterwordstretchfactor\fontdimen3\font minus
  \fontdimen4\font\relax}
\providecommand{\BIBforeignlanguage}[2]{{%
\expandafter\ifx\csname l@#1\endcsname\relax
\typeout{** WARNING: IEEEtran.bst: No hyphenation pattern has been}%
\typeout{** loaded for the language `#1'. Using the pattern for}%
\typeout{** the default language instead.}%
\else
\language=\csname l@#1\endcsname
\fi
#2}}
\providecommand{\BIBdecl}{\relax}
\BIBdecl

\bibitem{Haykin2005}
S.~Haykin, ``Cognitive radio: brain-empowered wireless communications,''
  \emph{IEEE J. Sel. Areas in Commun.}, vol.~23, no.~2, pp. 201--220, Feb 2005.

\bibitem{Goldsmith2009}
A.~Goldsmith, S.~Jafar, I.~Maric, and S.~Srinivasa, ``Breaking spectrum
  gridlock with cognitive radios: An information theoretic perspective,''
  \emph{Proc. IEEE}, vol.~97, no.~5, pp. 894--914, May 2009.

\bibitem{Madsen2005}
A.~Host-Madsen and J.~Zhang, ``Capacity bounds and power allocation for
  wireless relay channels,'' \emph{IEEE Trans. Inf. Theory}, vol.~51, no.~6,
  pp. 2020--2040, June 2005.

\bibitem{Bing2008}
B.~Xia, Y.~Fan, J.~Thompson, and H.~Poor, ``Buffering in a three-node relay
  network,'' \emph{IEEE Trans. on Wireles Commun.}, vol.~7, no.~11, pp.
  4492--4496, November 2008.

\bibitem{Zlatanov2011}
N.~Zlatanov, R.~Schober, and P.~Popovski, ``Throughput and diversity gain of
  buffer-aided relaying,'' in \emph{Proc. IEEE GLOBECOM, Houston,TX, USA}, Dec
  2011, pp. 1--6.

\bibitem{Zlatanov2013_2}
N.~Zlatanov and R.~Schober, ``Buffer-aided relaying with adaptive link
  selection---fixed and mixed rate transmission,'' \emph{IEEE Trans. Inf.
  Theory}, vol.~59, no.~5, pp. 2816--2840, May 2013.

\bibitem{Zlatanov2013_1}
N.~Zlatanov, R.~Schober, and P.~Popovski, ``Buffer-aided relaying with adaptive
  link selection,'' \emph{IEEE J. Sel. Areas in Commun.}, vol.~31, no.~8, pp.
  1530--1542, August 2013.

\bibitem{Islam2014}
T.~Islam, D.~Michalopoulos, R.~Schober, and V.~K. Bhargava, ``Delay constrained
  buffer-aided relaying with outdated csi,'' in \emph{Proc. IEEE WCNC,
  Istanbul, Turkey}, April 2014, pp. 875--880.

\bibitem{Kim2013}
K.~J. Kim, T.~Duong, and H.~Poor, ``Outage probability of single-carrier
  cooperative spectrum sharing systems with decode-and-forward relaying and
  selection combining,'' \emph{IEEE Trans. Wireles Commun.}, vol.~12, no.~2,
  pp. 806--817, February 2013.

\bibitem{Darabi2014}
M.~Darabi, B.~Maham, X.~Zhou, and W.~Saad, ``Buffer-aided relay selection with
  interference cancellation and secondary power minimization for cognitive
  radio networks,'' in \emph{Proc. IEEE DYSPAN, Mclean, VA, USA}, April 2014,
  pp. 137--140.

\bibitem{Darabi2015}
M.~Darabi, V.~Jamali, B.~Maham, and R.~Schober, ``Adaptive link selection for
  cognitive buffer-aided relay networks,'' \emph{IEEE Commun. Lett.}, vol.~19,
  no.~4, pp. 693--696, April 2015.

\bibitem{Chen2014}
G.~Chen, Z.~Tian, Y.~Gong, and J.~Chambers, ``Decode-and-forward buffer-aided
  relay selection in cognitive relay networks,'' \emph{IEEE Trans. on Veh.
  Technol.}, vol.~63, no.~9, pp. 4723--4728, Nov 2014.

\bibitem{Shaqfeh2015}
M.~Shaqfeh, A.~Zafar, H.~Alnuweiri, and M.~Alouini, ``Overlay cognitive radios
  with channel-aware adaptive link selection and buffer-aided relaying,''
  \emph{IEEE Trans. Commun.}, vol.~PP, no.~99, pp. 1--1, 2015.

\bibitem{Hong2014}
X.~Hong, J.~Wang, C.-X. Wang, and J.~Shi, ``Cognitive radio in 5g: a
  perspective on energy-spectral efficiency trade-off,'' \emph{IEEE Commun.
  Magazine}, vol.~52, no.~7, pp. 46--53, July 2014.

\bibitem{Haider2015}
F.~Haider, C.-X. Wang, H.~Haas, E.~Heps., X.~Ge, and D.~Yuan, ``Spectral and
  energy efficiency analysis for cognitive radio networks,'' \emph{IEEE Trans.
  Wireless Commun.}, vol.~14, no.~6, pp. 2969--2980, June 2015.

\bibitem{Lee2011}
J.~Lee, H.~Wang, J.~Andrews, and D.~Hong, ``Outage probability of cognitive
  relay networks with interference constraints,'' \emph{IEEE Trans. Wireles
  Commun.}, vol.~10, no.~2, pp. 390--395, February 2011.

\bibitem{Abramowitz1965}
M.~Abramowitz and I.~Stegun, \emph{Handbook of Mathematical Functions}.\hskip
  1em plus 0.5em minus 0.4em\relax Dover Publications, 1965.

\end{thebibliography}
\end{document}